\documentclass[fleqn,10pt]{wlscirep}
\usepackage[utf8]{inputenc}
\usepackage[T1]{fontenc}
\title{Emerging solutions from the battle of defensive alliances}

\author[1,*]{Attila Szolnoki}
\author[2]{Xiaojie Chen}
\affil[1]{Institute of Technical Physics and Materials Science, Centre for Energy Research, P.O. Box 49, H-1525 Budapest, Hungary}
\affil[2]{School of Mathematical Sciences, University of Electronic Science and Technology of China, Chengdu 611731, China}

\affil[*]{Correspondence to szolnoki.attila@ek-cer.hu}

\begin{abstract}
Competing strategies in an evolutionary game model, or species in a biosystem, can easily form a larger unit which protects them from the invasion of an external actor. Such a defensive alliance may have two, three, four or even more members. But how effective can be such formation against an alternative group composed by other competitors? To address this question we study a minimal model where a two-member and a four-member alliances fight in a symmetric and balanced way. By presenting representative phase diagrams, we systematically explore the whole parameter range which characterizes the inner dynamics of the alliances and the intensity of their interactions. The group formed by a pair, who can exchange their neighboring positions, prevail in the majority of the parameter region. The rival quartet can only win if their inner cyclic invasion rate is significant while the mixing rate of the pair is extremely low. At specific parameter values, when neither of the alliances is strong enough, new four-member solutions emerge where a rock-paper-scissors-like trio is extended by the other member of the pair. These new solutions coexist hence all six competitors can survive. The evolutionary process is accompanied by serious finite-size effects which can be mitigated by appropriately chosen prepared initial states.  
\end{abstract}
\begin{document}

\flushbottom
\maketitle

\thispagestyle{empty}

\section*{Introduction}

If I am stronger than the enemy of my neutral partner then the mentioned intruder can be blocked. Similarly, my own enemy may hopefully be beaten by my neutral partner. The only crucial point is neutral partners should be capable to exchange their positions hence giving a chance for the proper defender to meet with the external intruder. This mechanism establishes the simplest two-member defensive alliance when biological species, or strategies in an evolutionary game model, compete for space \cite{szabo_pre07,szolnoki_amc23}. Practically similar indirect collaboration can also emerge in larger groups where members are not neutral, but they are in a predator-prey relationship to each other \cite{szabo_jpa05,brown_pre19}. The latter formations are based on nonhierarchical, or in other words, intransitive competition of members \cite{avelino_pre19b,tainaka_ei21,park_c22,yoshida_srep22}. The simplest version is when the relation of three actors can be characterized by the well-known rock-scissors-paper game \cite{szolnoki_jrsif14,mobilia_g16,avelino_epl21,nagatani_jtb19,szolnoki_pre16,avelino_epl18}. While the members are basically enemies, but they can protect indirectly each other from an external invader. It is worth stressing that the mentioned cyclic dominance can be observed in very different biological systems, including bacterias, plants, or animals \cite{kerr_n02,garde_rsob20,cameron_jecol09,ruifrok_tpb15,szolnoki_srep16b,burrows_mep98}. 

For simplicity, in the remaining of this paper we refer the actors as species, but our abstract model could be valid in other research areas, too. For example, cyclic dominance, or intransitive relation between actors can also be detected when different strategies compete in an evolutionary game model \cite{szolnoki_njp14,szolnoki_csf20b,roman_jsm12}. A simple example is when cooperator, defector and loner strategies beat each other cyclically, hence providing a stable coexistence for all competitors \cite{hauert_s02}. Further examples of cyclic relation were also found when more sophisticated strategies, like conditional cooperators, or so-called informed strategies are present \cite{szolnoki_epl15}. Evidently, larger loops with four-, five-, six-, or even more species are also possible to form more subtle alliances \cite{esmaeili_pre18,intoy_jsm13,park_c19b,vukov_pre13,baker_jtb20,park_c19c,szabo_pre08,roman_jtb16,bayliss_pd20,park_csf23,menezes_csf23}.

The fight can step onto a higher level when the external invader is not a single species, but also an alternative alliance. In this case the inner dynamics of an alliance could be a decisive factor to beat the alternative group. For instance, when two three-member groups compete then the faster inner invasion makes a cyclic loop fitter than the other trio where invasions are less intensive \cite{perc_pre07b}. It could also be crucial how homogeneous the invasion rates within a group, because highly heterogeneous rates make a trio vulnerable no matter the average invasion in the loop is high \cite{szolnoki_epl20}. The conclusions could be more complicated when alliances with different sizes compete. The simplest symmetrical model could be when a two-member and a four-member groups fight in a six-species system. In this work we study the symmetric model of six species where the relation of all actor can be described by three parameters. The first parameter characterizes the intensity of cyclic invasion among the four-member group, while the second parameter determines the frequency of neighboring site exchange of the pair. Last, the third parameter decides the intensity of interaction between the two competing alliances. Our principal goal is to explore the complete parameter space and find the possible winner at each parameter combinations. In this way we can identify principles which generally determine the vitality of an alliance when competing formations are also on the stage.

\section*{Model}

Defensive alliances require a spatial setting of competing actors, therefore in our minimal model the species are distributed on a square lattice where each lattice point is occupied by one of the $s_i$ $i \in \{\it 1, 2, 3, 4, 5, 6\}$ six species. Periodic boundary conditions are applied, hence all players can interact with one of the four nearest neighbors. By forming a four-member loop, species {\it 1, 2, 3} and {\it 4} invade cyclically each other with probability $\alpha$. The other group is formed by species {\it 5} and {\it 6} who are neutral, but they exchange their neighboring position with probability $\beta$. Last, the two alliances invade each other in a balanced way. More precisely, species {\it 1} and species {\it 3} invades species {\it 5}, while the latter invades species {\it 2} and species {\it 4}. Similarly, species {\it 2} and species {\it 4} invades species {\it 6}, while the latter invades species {\it 1} and species {\it 3}. The invasions between the alliances happen with probability $\gamma$. Summing up, the microscopic rules are defined in the following way:
\begin{align}
	&s_i s_{i+1} \xrightarrow{\alpha} s_i s_i \,\,\,{\rm if} \,\,\, i \in \{1, 2, 3, 4\} \,\,\,{\rm in \,\,\, a\,\,\, cyclic \,\,\, manner} \nonumber\\
	&s_5 s_6 \xrightarrow{\beta} s_6 s_5 \label{rules}\\
	&s_1 s_5 \xrightarrow{\gamma} s_1 s_1, \,s_3 s_5 \xrightarrow{\gamma} s_3 s_3, \,s_2 s_6 \xrightarrow{\gamma} s_2 s_2, \,s_4 s_6 \xrightarrow{\gamma} s_4 s_4, \,\,\,{\rm and} \,\,\,
	s_5 s_2 \xrightarrow{\gamma} s_5 s_5, \,s_5 s_4 \xrightarrow{\gamma} s_5 s_5, \,s_6 s_1 \xrightarrow{\gamma} s_6 s_6, \,s_6 s_3 \xrightarrow{\gamma} s_6 s_6 \,.
	\nonumber
\end{align}

\begin{figure}[ht]
\centering
\includegraphics[width=9cm]{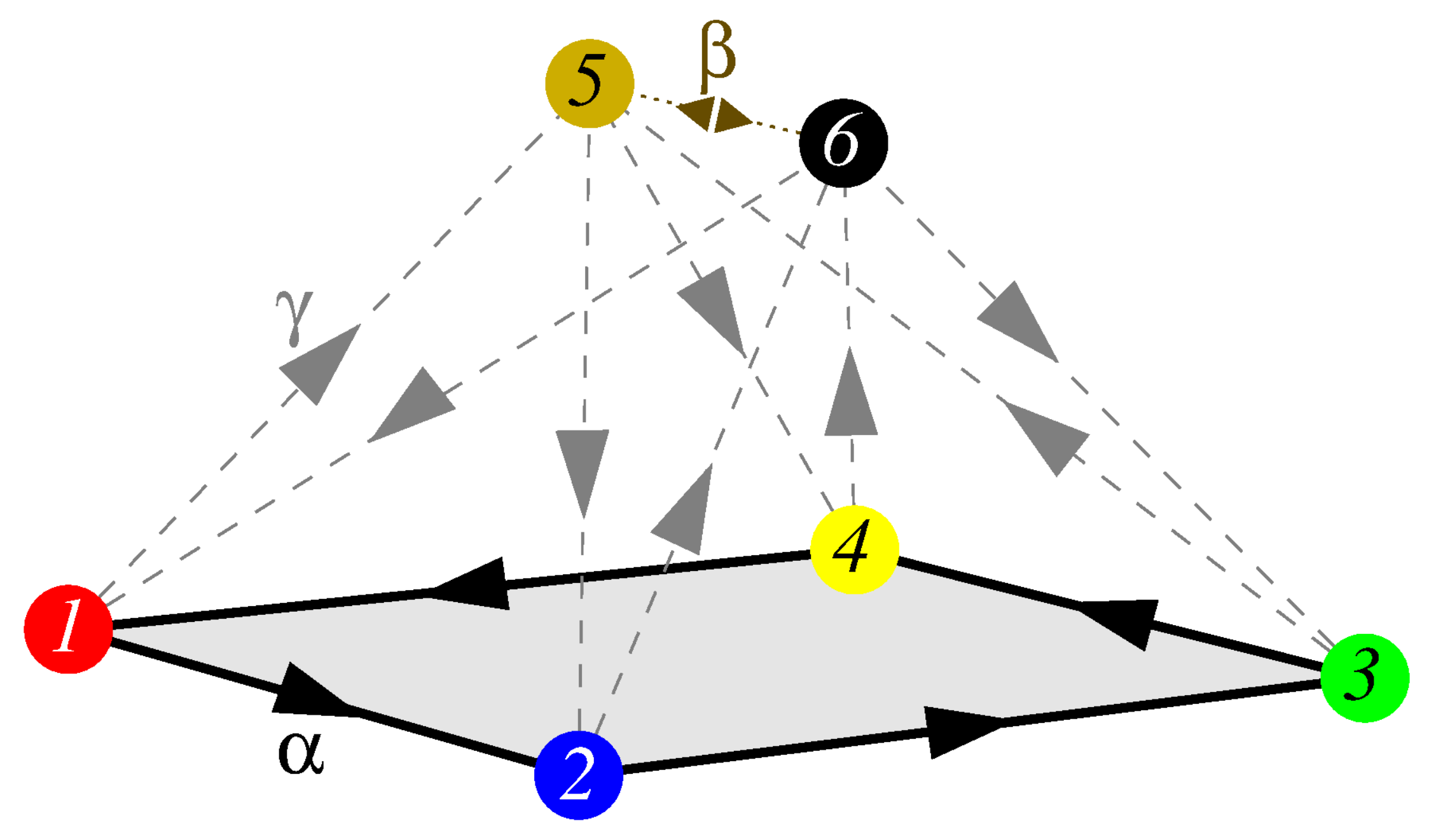}
\caption{Food-web of the minimal symmetric model where a two-member and a four-member alliances meet. Solid black arrows denote the cyclic invasion among species {\it 1, 2, 3} and {\it 4} with probability $\alpha$. Species $\it 5$ and $\it 6$ can exchange their positions with probability $\beta$. The mutual invasion between the mentioned groups happens with probability $\gamma$, as indicated by the dashed grey arrows.}
\label{foodweb}
\end{figure}

To give a clear description about the relation of the species we present the food-web of all participants in Fig.~\ref{foodweb}. This graph highlights that parameter $\alpha$ and parameter $\beta$ characterize the inner dynamics of the alliances while parameter $\gamma$ determines the interaction strength between these formations. Importantly, the relation between these groups are balanced, namely species {\it 5} is a predator of two members of the quartet and simultaneously the prey of the remaining two members of the loop. Similar can be said about the relation of species {\it 6} and the quartet. It is also worth noting that in the absence of species {\it 6} the external invader species {\it 5} would be crowed out by the cyclic loop of the quartet. Naturally, alone species {\it 6} would also be vulnerable against the four-member formation. This is a well-known mechanism how a cyclic loop is capable to defense an external intruder. But now, both species {\it 5} and species {\it 6} are present simultaneously, and more importantly, they can exchange their positions. Our main goal is to reveal the effectiveness of their defensive alliance when they fight against another formation.

At this point it is important to note that a conceptually similar system was studied recently \cite{de-oliveira_csf22}. Our present model, however, posses the most symmetric setup of possible interactions between the competing alliances. In this way the conditions for a fair competition are established. As a consequence, four additional three-member loops emerge in the food-web due to these external interactions. For example, species {\it 1}, {\it 2}, and {\it 6} form a rock-scissors-paper-like group, but there are three similar formations, including {\it 2+3+5}, {\it 3+4+6}, and {\it 1+5+4}. They are selected from the members of the original competing alliances. As we will show later, these possibilities have a serious consequence on evolution and give a chance for new solutions to emerge.

Monte Carlo (MC) simulations are carried out in the full three-dimensional $(\alpha,\beta,\gamma)$ parameter space. During an elementary step we randomly choose a node and a nearest neighbor in the $L \times L$ lattice. If they are occupied by different species then we execute the possible change with the given probability defined by Eqs.~\ref{rules}. A full MC step contains $L \times L$ elementary steps. We need to vary the linear system size between $L=400$ and $L=3600$ to avoid serious finite size effects in the vicinity of phase transition points. It is also important to stress that the traditionally used random initial state, when species are distributed randomly when we launch the evolution, does not always give reliable information about the stationary state which is valid in the large-size limit. Therefore we also use alternative initial states where the evolution starts from a prepared initial state \cite{szolnoki_pre11b,serrao_epjb21,szolnoki_njp16,mir_pre22}. The details are given in the next section. In general, the solution which is considered as a stable solution, should be obtained starting from different initial states where all competing species are present. Naturally, the relaxation time to reach the mentioned state or the minimal system size could be very different and it may also depend on how far we are from a specific phase transition point. In our present model the necessary relaxation time is between $10^4$ to $2 \cdot 10^6$ MC steps.

\section*{Results}

\subsection*{Strong interaction between alliances}

\begin{figure}[ht]
	\centering
	\includegraphics[width=9cm]{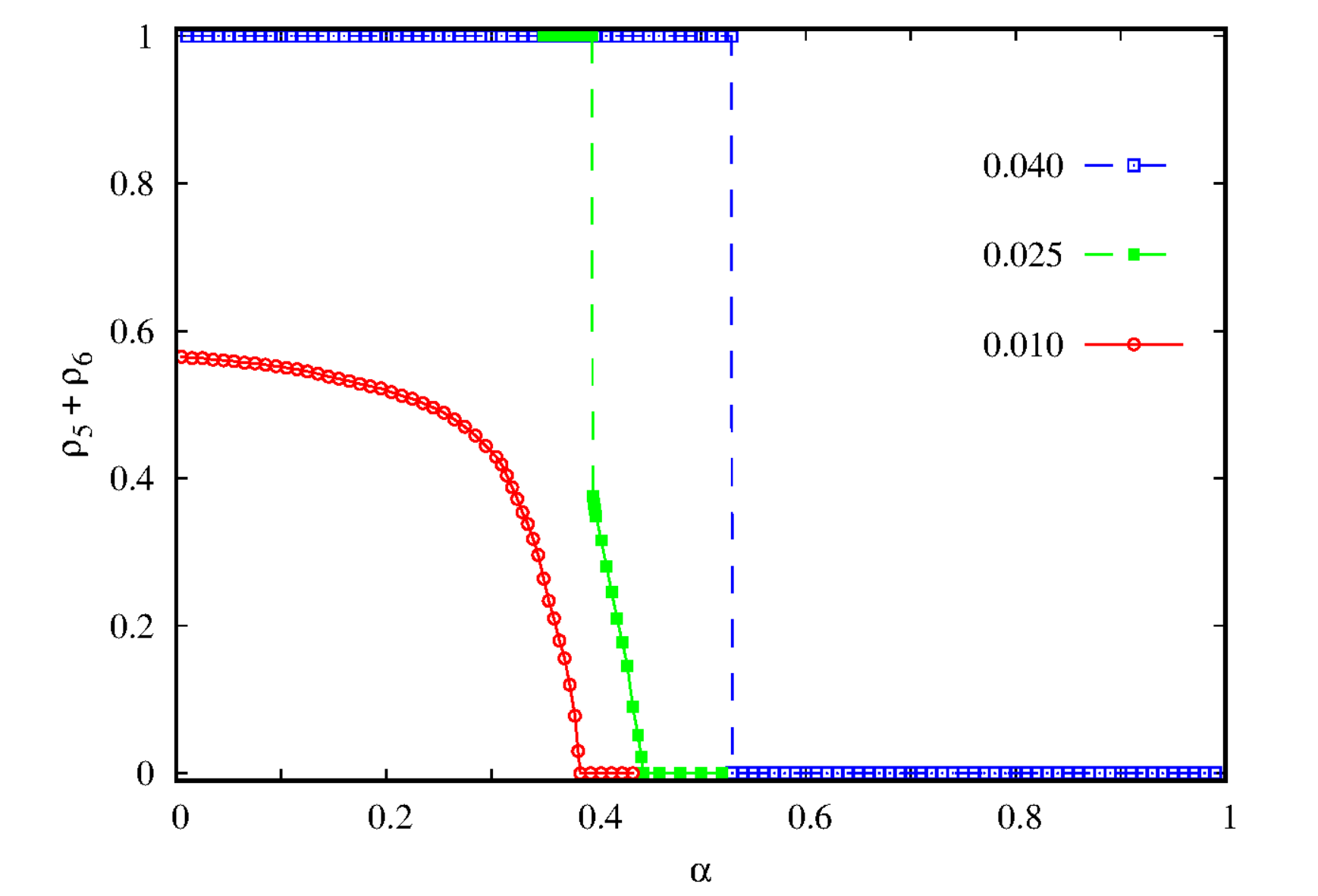}\includegraphics[width=7.7cm]{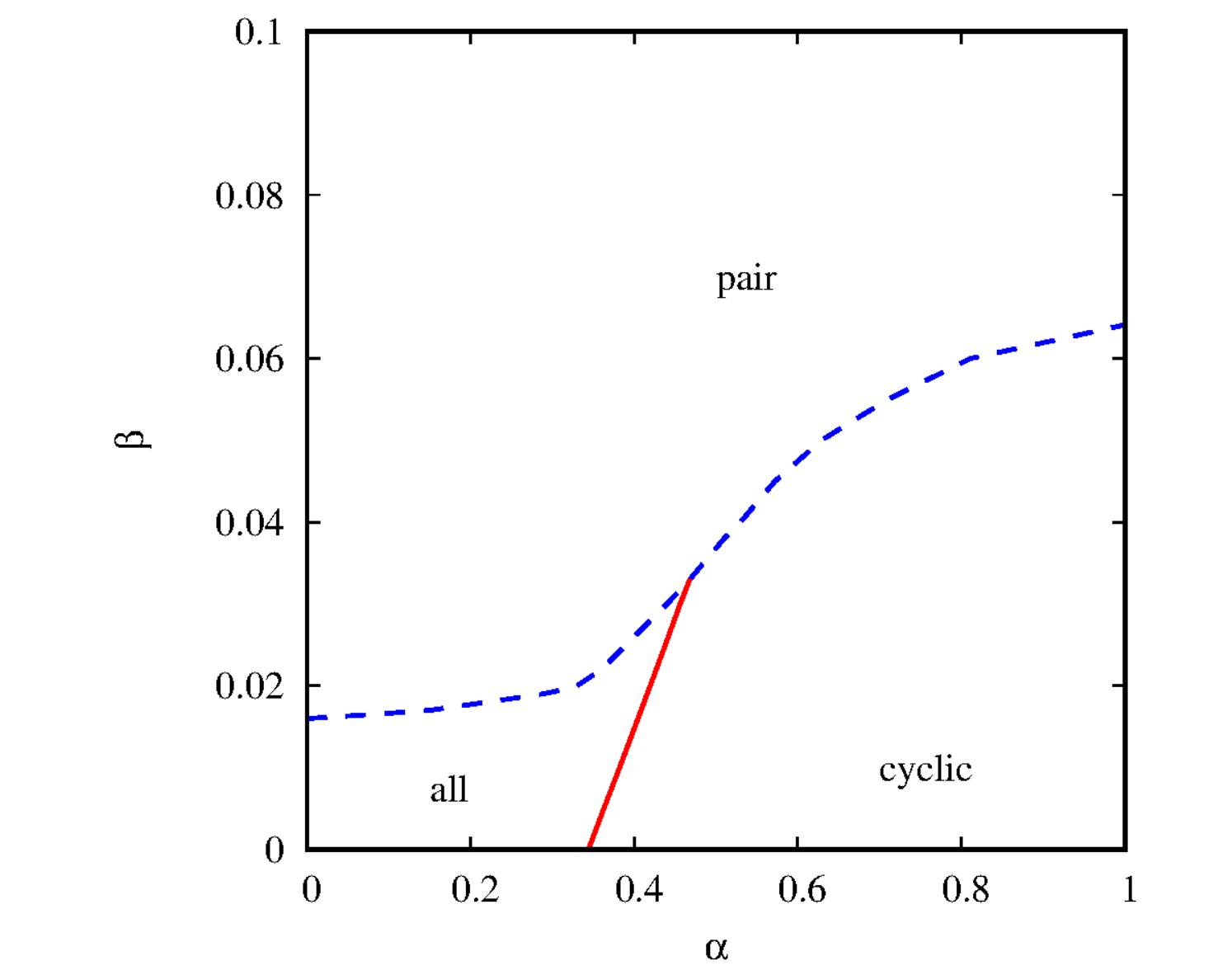}
	\caption{Representative system behavior at $\gamma=0.9$ when the interaction between the competing alliances is strong. Left panel depicts how the order parameter changes as we vary the value of $\alpha$ for three representative mixing rates. The values of $\beta$ are shown in the legend. The order parameter, which characterizes the stable solutions, is the $\rho_5+\rho_6$ sum of the portions of the pair species. For $\beta=0.04$ the transition is discontinuous, while for $\beta=0.01$ we can see a continuous transition. For intermediate $\beta=0.025$ a discontinuous  transition is followed by a continuous one. The right panel summarizes the system behavior on the $\beta-\alpha$ parameter plane. Here ``all'' marks the phase when all six competing strategies survive. Red solid line marks the positions of continuous, while blue dashed line denotes the location of discontinuous transition points.}
	\label{g9}
\end{figure}

We first present the system behavior in the large $\gamma$ region when the interaction between the pair and the quartet alliances is intensive. If we fix the value of $\gamma$ then two free parameters remain which determine the inner dynamics of the competing alliances. If the $\beta$ mixing rate between species {\it 5} and species {\it 6} is high then their couple prevail independently of the value $\alpha$. If we lower the value of $\beta$ then the alternative alliance can win for large $\alpha$ values. The appropriate order parameter, which characterizes the transition between the mentioned solutions, is the sum of $\rho_5$ and $\rho_6$ portions of the pair species. Evidently, this sum is equal to 1 in the ``pair'' solution and becomes zero in the ``cyclic'' solution. The left panel of Fig.~\ref{g9} illustrates that the ``pair'' solution is replaced by cyclic solution of the quartet at $\alpha=0.529$ via a discontinuous transition if $\beta=0.04$. For smaller mixing rate, as it is shown for $\beta=0.025$, a new phase emerges for intermediate $\alpha$ values. In this case all six species coexist. If we increase the value of $\alpha$ further then this solution gradually disappears and is replaced by the cyclic solution via a continuous phase transition. Last, if we further decrease the $\beta$ value then the ``pair'' solution completely disappears. 

The complete behavior is summarized in the phase diagram shown on the right panel of Fig.~\ref{g9} where the stable solutions are depicted on the $\beta-\alpha$ parameter plane. We note that a relatively small mixing rate is enough for the pair alliance to beat all other solutions. More precisely, if $\beta > 0.07$ then the mixture of species {\it 5} and species {\it 6} prevail even if the invasion rate among the quartet species is maximal. The latter can only win in the right-down corner of the parameter plane when $\alpha$ is high, hence the quartet is strong, and $\beta$ is small, hence the pair alliance is weak. The left-down corner of the diagram represents the case when both the pair and the cyclic alliances are weak. Here a new solution emerges where all six species are present. Later, we will describe the features of this new solution in detail.

\begin{figure}[ht]
	\centering
	\includegraphics[width=9cm]{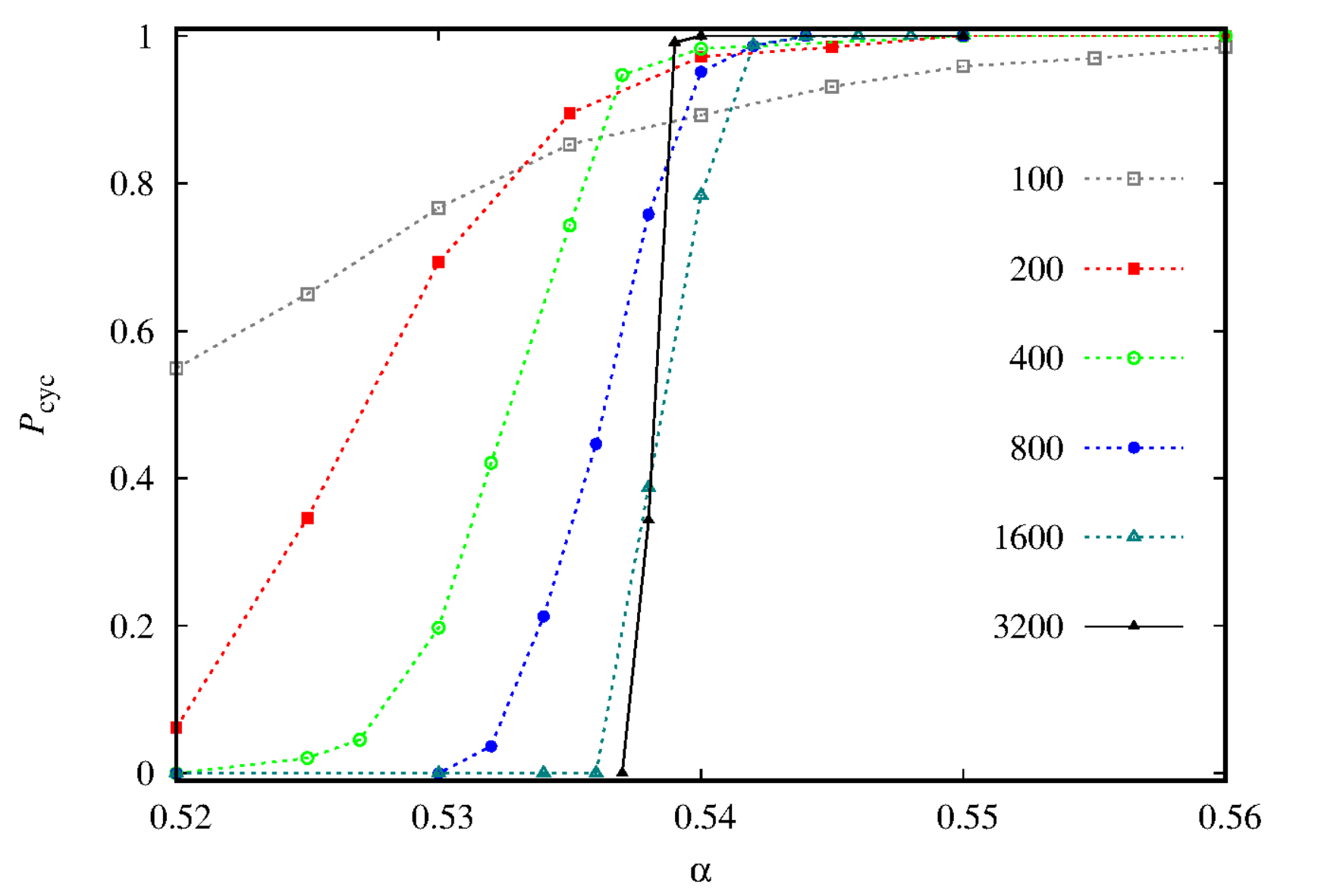}\includegraphics[width=6cm]{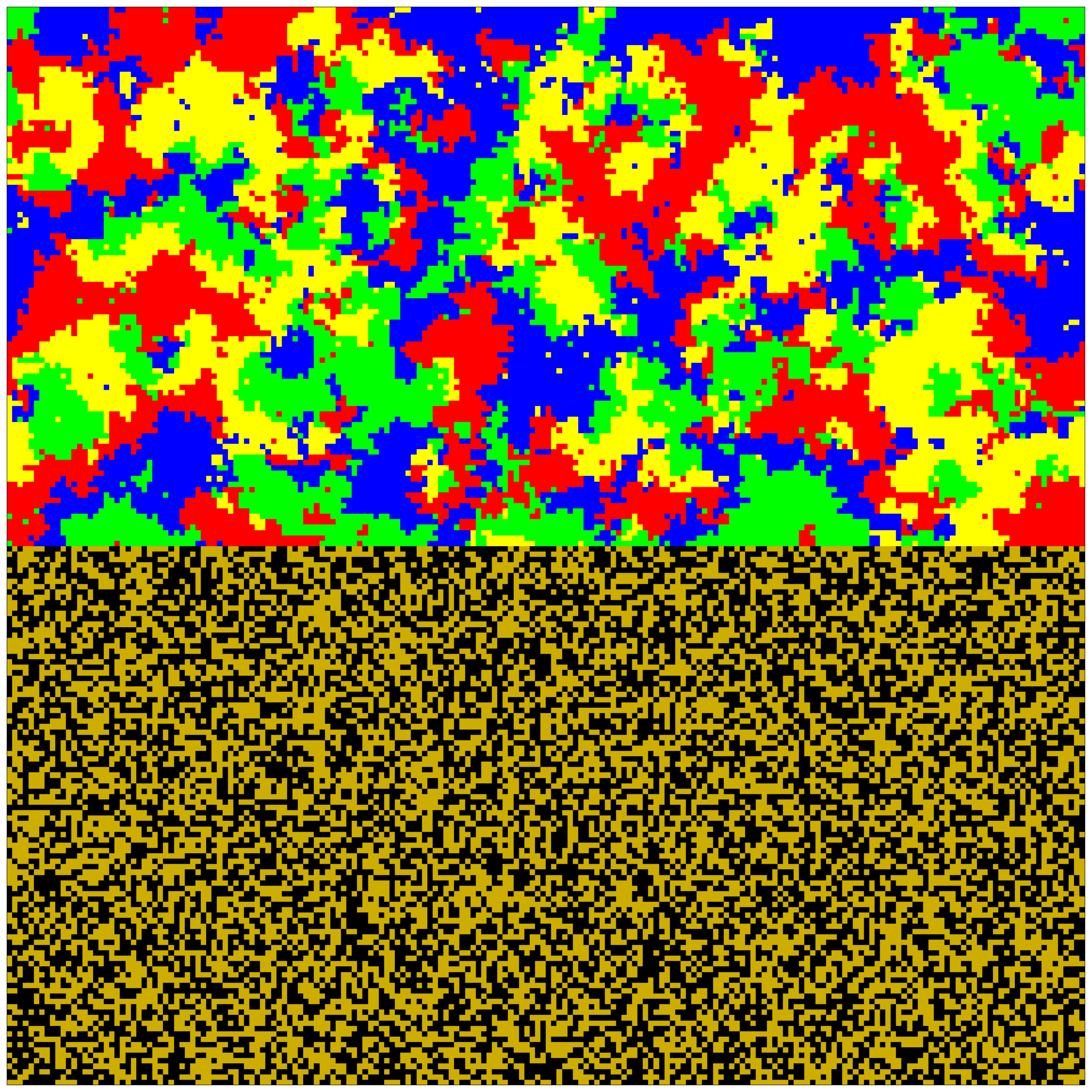}
	\caption{The left panel shows the fixation probability to the state where the cyclic four-species alliance prevails in dependence of $\alpha$ at fixed $\beta=0.04$. The initial state is always a random distribution of all six species. In case of the alternative destination the system evolves onto the ``pair'' solution. The linear system sizes are indicated in the legend. The right panel illustrates a prepared initial state, obtained at $L=200$, where both alliances are ready to compete from the very beginning. The usage of this initial state mitigates the finite-size effects significantly. The colors to represent species agree with the code we used in Fig.~\ref{foodweb}.}
	\label{fs}
\end{figure}

But first, let us briefly comment the technical difficulties of identifying the phase transition points in this system. It is a standard protocol that we start the simulation from an initial state where the spatial distribution of competing species is random. If we use such a starting state then we can observe a serious finite-size effect now. More precisely, the evolutionary outcome is ambiguous, sometimes the system terminates into the ``pair'' solution, but the alternative final state, where the cyclic quartet remains, may also be observed at the same parameter values. To illustrate this uncertainty, in the left panel of Fig.~\ref{fs} we present a statistics about the possible destinations. More precisely, we plot the probability to reach the cyclic four-strategy state for different $\alpha$ values at a fixed $\beta$. Importantly, the alternative evolutionary outcome is to reach the ``pair'' state. To obtain these numbers we have executed 200-2000 independent runs depending on the lattice size at each $\alpha$ values. The plot shows clearly that we cannot really trust on the numerical results obtained for smaller system sizes. For instance, at $\alpha=0.51$, which is quite far from the proper transition point, both possible destinations are almost equally likely for $L=100$ system size. The transition point, where ``pair'' solution is replaced by the alternative cyclic alliance, is at $\alpha=0.538(2)$. But we note that the outcome is still ambiguous for $L=3200$. Importantly, this finite-size problem can be mitigated significantly if we use an alternative initial state where both alliances are present from the very beginning and they can compare their vitality in a fair way. The right panel of Fig.~\ref{fs} shows such an initial state where we divided the available space into two halves and each sector is occupied by one of the alliances. In this way the the possible solutions can compete properly, hence the final evolutionary outcome is less ambiguous. In particular, at the previously mentioned parameters we can already obtain reliable data for the location of the phase transition point by using $L=800$ linear system size.

\begin{figure}[ht]
	\centering
	\includegraphics[width=17cm]{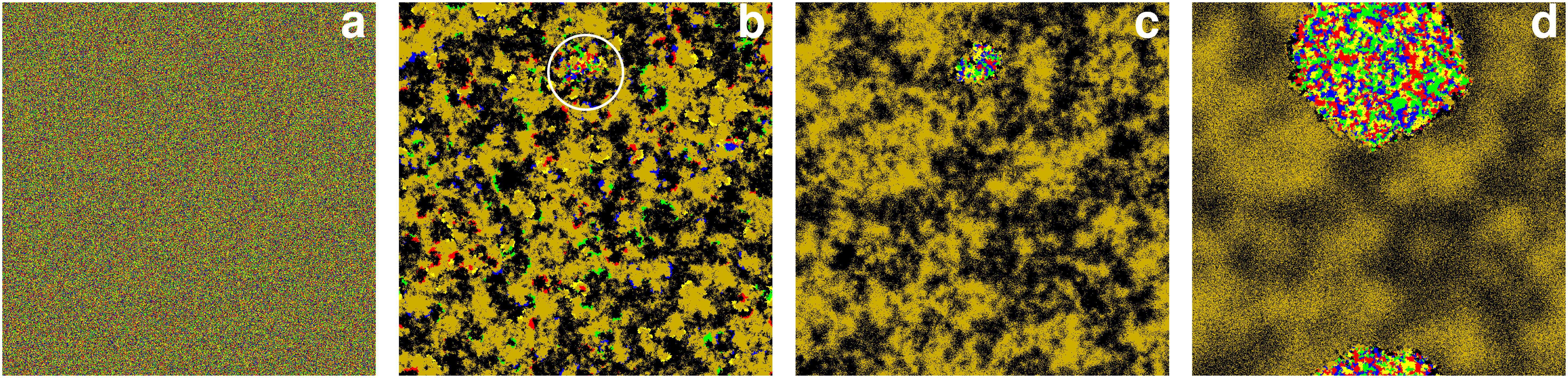}
	\caption{We start the evolution from a random initial state, as shown in panel~(a). After 800 MC steps, shown in panel~(b), the cyclic alliance emerges. The seed of this solution is marked by a white circle. The remaining part of available space is conquered by the ``pair'' solution, as illustrated by panel~(c), taken after 2200 MC steps. Slowly, but surely the cyclic alliances invades the rival group. An intermediate stage of the process, obtained after 50,000 MC steps, is shown in panel~(d). Finally, not shown here, the four-member group prevails. Other parameters are $\alpha=0.9$, $\beta=0.08$, $\gamma=0.2$, $L=800$. The color code to designate species is the same we used earlier.}
	\label{seed}
\end{figure}

To understand the origin of the evolutionary uncertainty, we monitor a representative pattern formation in Fig.~\ref{seed}. Initially, all six species are randomly distributed as shown in panel~(a). When we launch the evolution then the pair of species {\it 5} and species {\it 6} can easily form their alliance which becomes effective against a randomized pattern. As panel~(b) illustrates, they gradually crowd out the remaining four species who try to fight against them separately. Just a single seed emerges where all members of the cyclic solution are present simultaneously, hence they can form the alternative alliance. This moment is shown by a white circle in panel~(b). In the absence of this formation, the {\it 5+6} pair would win, as it is demonstrated in panel~(c) where the majority of available space is occupied by this solution. The rival alliance, however, can resist, and more importantly, they can reverse the direction of the war and gradually invade the whole space. Notably, this invasion could be a slow process. For example, panel~(d) of Fig.~\ref{seed} was taken after 50,000 MC steps, but the final destination to reach the full cyclic solution is unavoidable and was recorded after 300,000 MC steps.

The presented example highlights the difficulty when we look for the winning solution. As we pointed out, it could be misleading to start the evolution from a random state because this setup does not necessarily give equal chance for all possible solutions. Importantly, to produce Fig.~\ref{seed} we used relatively large system size, where $L=800$, still, the chance of emerging the proper winner solutions was low. For example, if we zoom a smaller area having $100 \times 100$, or even $400 \times 400$ size then the system can easily evolve to the ``pair'' solution. Of course, the ``cyclic'' solution can also be reached at smaller size, if we are lucky enough to observe the surviving seed of the mentioned solution. This explains the ambiguity at smaller system sizes. Evidently, as we increase the system size, the mentioned uncertainty becomes less likely because somewhere in a larger population there is always a chance for cyclic solution to emerge and test its vitality against the pair solution. But in some cases, depending on the parameter values, we may need unusually large system size to give equal chance for all competing solutions. Just to illustrate the degree of the problem, at certain parameter values, even an $L=3200$ linear system size can produce hectic destination. Luckily, this problem can be practically avoided if we apply a prepared initial state where both competing alliances are present from the very beginning hence their competition gives a more reliable information about the proper winner.

\begin{figure}[ht]
	\centering
	\includegraphics[width=5cm]{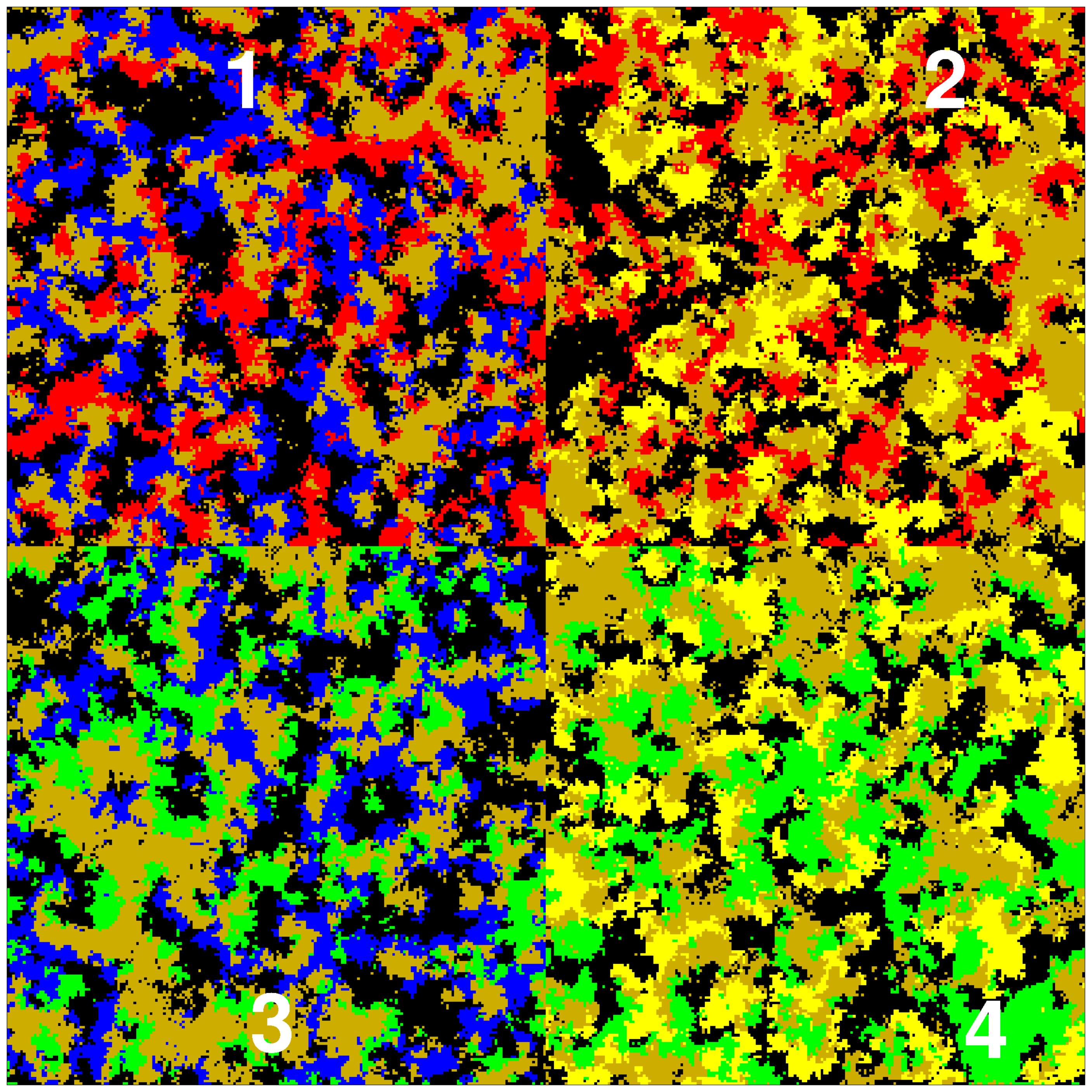}\hspace{3cm}\includegraphics[width=5cm]{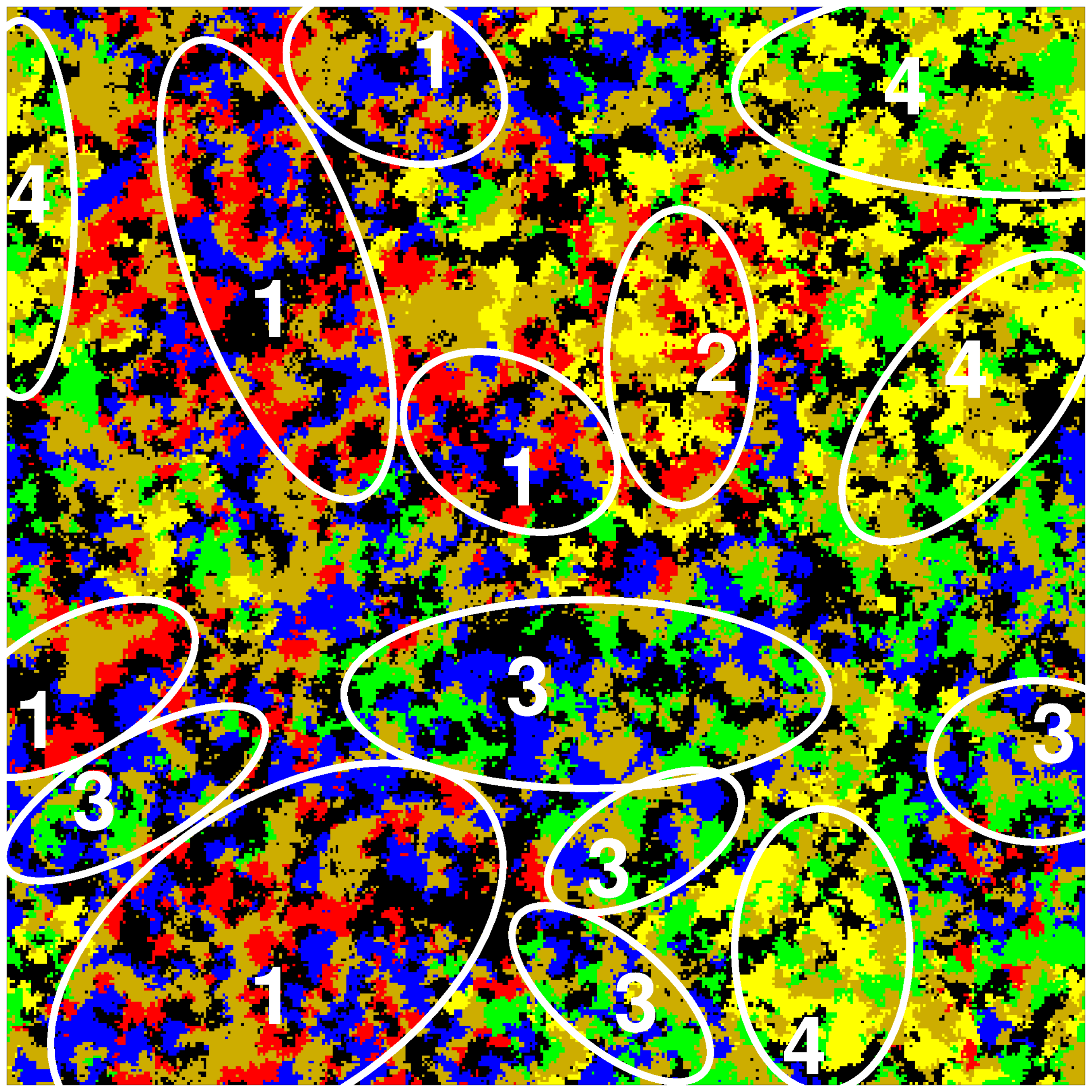}
	\caption{The left panel shows the possible four-member solutions formed by species selected from both alliances. They are numbered for easier reference. Solution $\# 1$ is formed by species {\it 1+2+5+6},  solution $\#2$ by species {\it 1+4+5+6}, solution $\# 3$ by {\it 2+3+5+6}, and solution $\# 4$ by {\it 3+4+5+6}. The right panel depicts a representative pattern of the ``all'' phase where all six species are present. As we marked by white ellipses, the pattern can be considered as the mixture of four-species solutions specified on the left side. Parameters are $\alpha=0.1$, $\beta=0.01$, $\gamma=0.9$, $L=400$. We use the same color code to designate species on the lattice.}
	\label{F1}
\end{figure}

In the following we discuss the so-called ``all'' phase when all competing species coexist. As we pointed out, this solution emerges when neither the ``pair'' nor ``cyclic'' alliances are strong enough to dominate the system. In the related parameter space the small system size may have another consequence. In particular, when the system size is small then it may happen that two of the species die out, while the portions of remaining four species remain stable. These four-member solutions, however, are different from the ``cyclic'' solution. One of them contains species {\it 1}, {\it 2}, {\it 5}, and {\it 6}. If we check the food-web shown in Fig.~\ref{foodweb} then we can easily identify a three-member cyclic loop here, formed by species {\it 1}, {\it 2}, and {\it 6}. Normally, this alliance would easily destroy the external species {\it 5}. But the site exchange between species {\it 5} and {\it 6} saves the intruder. Similar quartet is formed by species {\it 2}, {\it 3}, {\it 5}, and {\it 6}. Furthermore, the loop of species {\it 3}, {\it 4}, and {\it 6} can also coexist with species {\it 5}. The last quartet is formed by species {\it 1}, {\it 4}, {\it 5}, and {\it 6}. As the left panel of Fig.~\ref{F1} illustrates, these combinations are forming stable solutions in the mentioned parameter region. The six-species solution can be considered as a mixture of these four-member solutions. A representative pattern of the stationary state is shown in the right panel of Fig.~\ref{F1}. The domains of these four-species solutions are constantly growing and shrinking hence giving chance for all six species to survive. 

In the mentioned panel we marked these domains and their interactions can also explain why the six-species solution could be fragile at small system size. More precisely, if the typical size of the domains is comparable to the linear size of the lattice then one types of the domains may go extinct. For example, solution $\# 2$ has the smallest portion on the right panel of Fig.~\ref{F1}. If this domain diminishes then the delicate balance between the four solutions is broken. This effect is strongly related to the original four-member loop between species {\it 1}, {\it 2}, {\it 3}, and {\it 4}. If solution $\# 2$ diminishes then the invasion activity between species {\it 1} and {\it 4} lowers significantly. As a typical reaction in a cyclic dominant system \cite{tainaka_pla95}, the decrease of invasion strength will indirectly increase the relative portions  of species {\it 2} and {\it 3}. In parallel, the extinction of solutions $\# 4$ and $\# 1$ also happen, hence the system terminates into the $\# 3$ solution. We stress, however, that the coexistence of the mentioned solutions are stable if the system size is large enough. Accordingly, the phase named ``all'' is a proper and stable composition of all six species, albeit its stability is based on the balance of four four-member solutions.

\subsection*{Weak interaction between alliances}

\begin{figure}[ht]
	\centering
	\includegraphics[width=9cm]{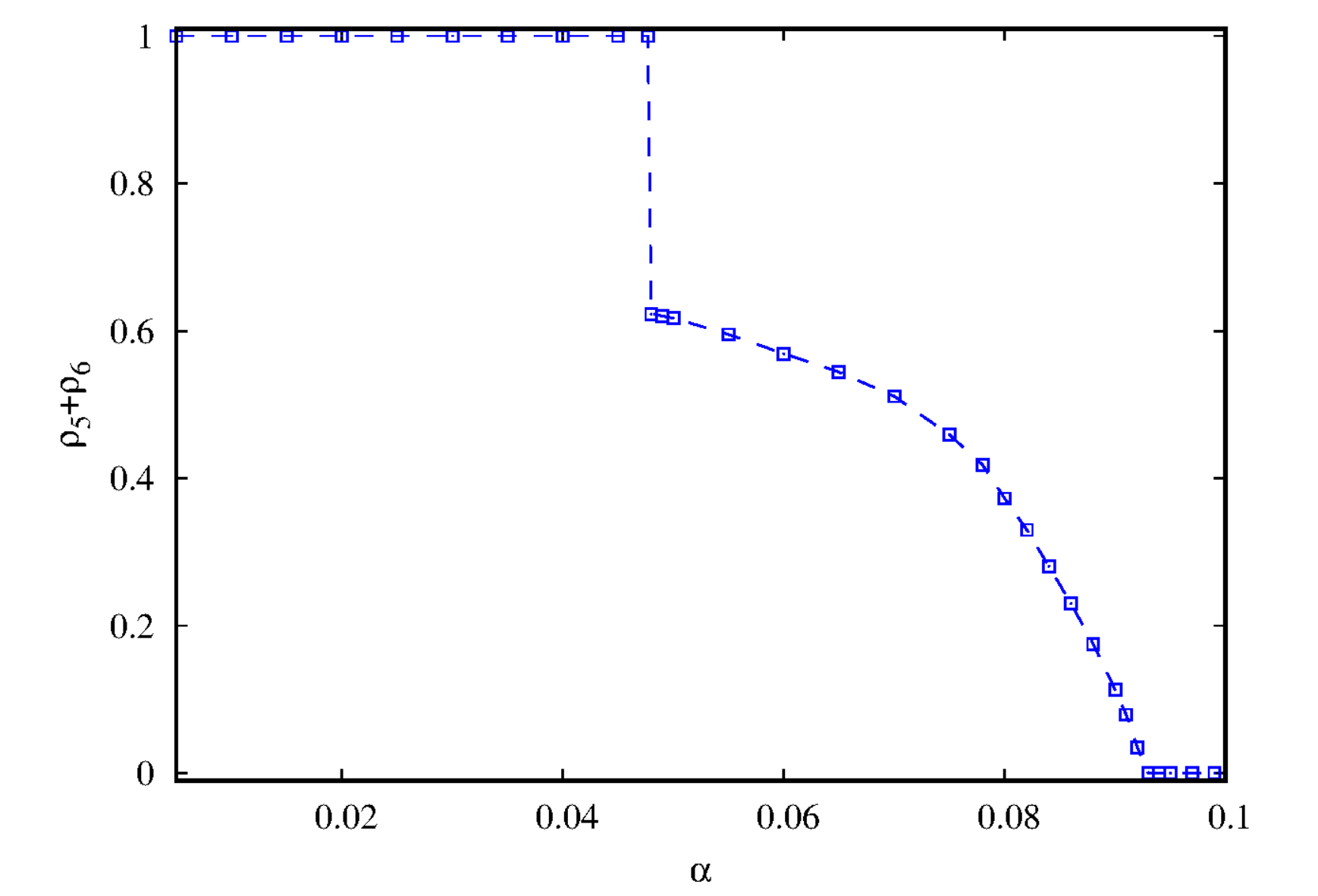}\includegraphics[width=7.7cm]{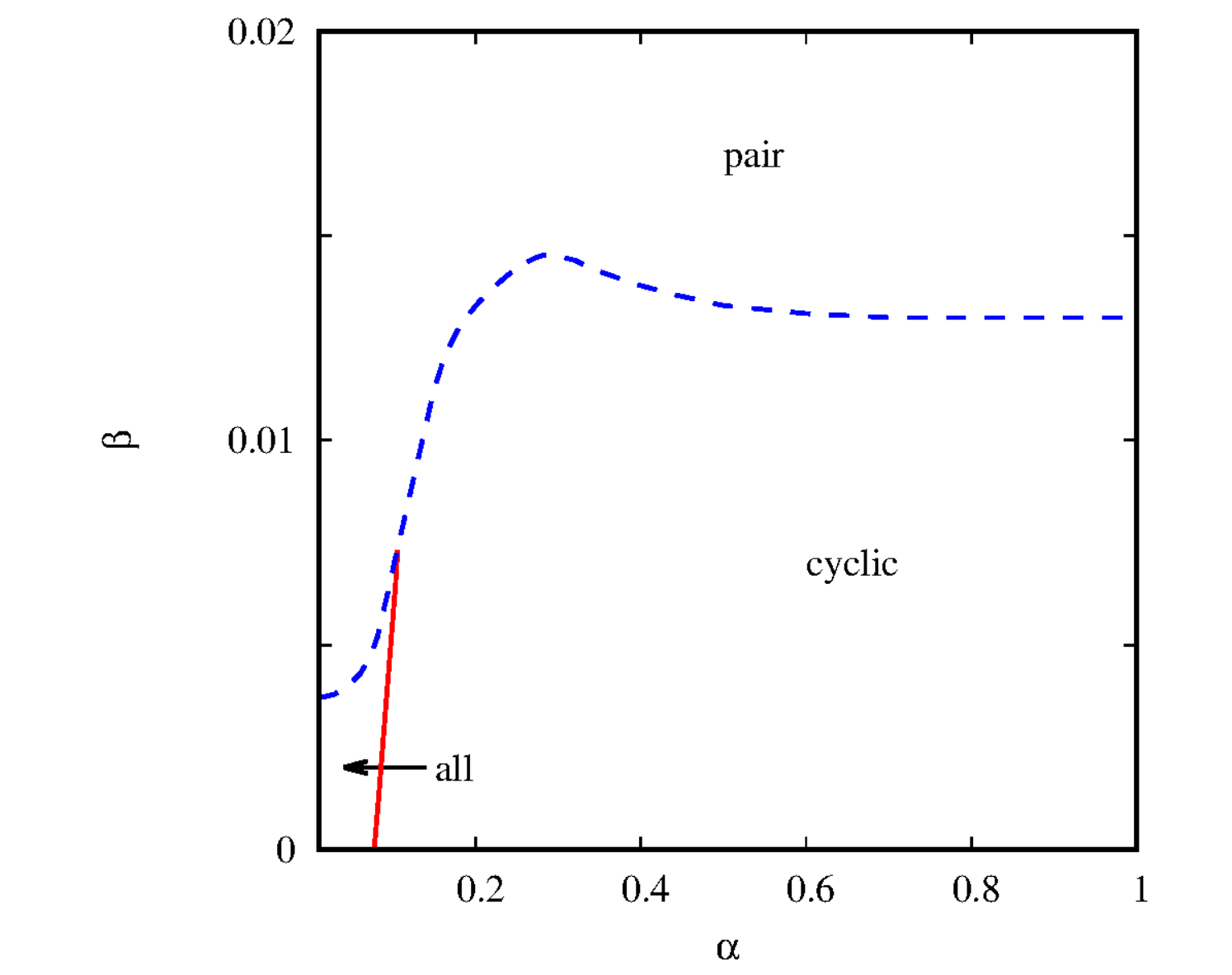}
	\caption{Right panel shows the phase diagram obtained at $\gamma=0.2$. The topology of the diagram is conceptually similar to the diagram shown inf Fig.~\ref{g9}. As previously, blue dashed (red solid) line marks the positions of discontinuous (continuous) phase transitions. Left panel shows a cross section obtained at $\beta=0.004$. As we increase $\alpha$, the system changes from ``pair'' to ``all'' to ``cyclic'' solution via a discontinuous and a continuous phase transition respectively. As previously, the order parameter is the sum of the portions of species {\it 5} and {\it 6}.}
	\label{g2}
\end{figure}

Next, we present the typical system behavior in the case when $\gamma$ is small, hence the interaction between the original alliances is weak. The resulting phase diagram is shown in the right panel of Fig.~\ref{g2}. We can see that the system behavior is basically similar to those we observed for strong interaction when $\gamma$ is large. Namely, if mixing rate between the pair species is large enough then they prevail and other solutions have no chance to survive. Furthermore, the rival cyclic alliance can only win if their inner invasion is intensive enough and $\beta$ is low. For small $\beta$ and $\alpha$ parameter pairs, however, the six-member solution can win. Similarly, the character of phase transitions is the same we reported for $\gamma=0.9$. An illustration is shown in the left panel of Fig.~\ref{g2} where the system arrives to ``all'' phase from ``pair'' solution and finally terminates to the ``cyclic'' phase as we increase $\alpha$. 

There is, however, a striking difference between the phase diagrams obtained for large and small $\gamma$ values. Namely, the kingdom of ``pair'' alliance is significantly larger for $\gamma=0.2$. Note that this solution is always dominant if $\beta > 0.015$ independently of the value of $\alpha$ which determines the power of cyclic alliance. For comparison, this threshold value is almost five times larger for $\gamma=0.9$. It simply means that the weak interaction between the competing alliances does not allow the cyclic solution to reveal their defending mechanism and the simpler site exchange dynamics is more efficient. The relatively small interaction between the alliances also makes the simulations difficult. Just to illustrate the difficulties, if we launch the evolution from a random initial state then even an $L=3200$ linear system size could be inadequate to identify the solution which is valid in the large system size limit. Even this linear system size can give different destinations at the same $\alpha-\beta$ parameter pairs if the evolution is started from a random configuration. 

\begin{figure}[ht]
	\centering
	\includegraphics[width=7.7cm]{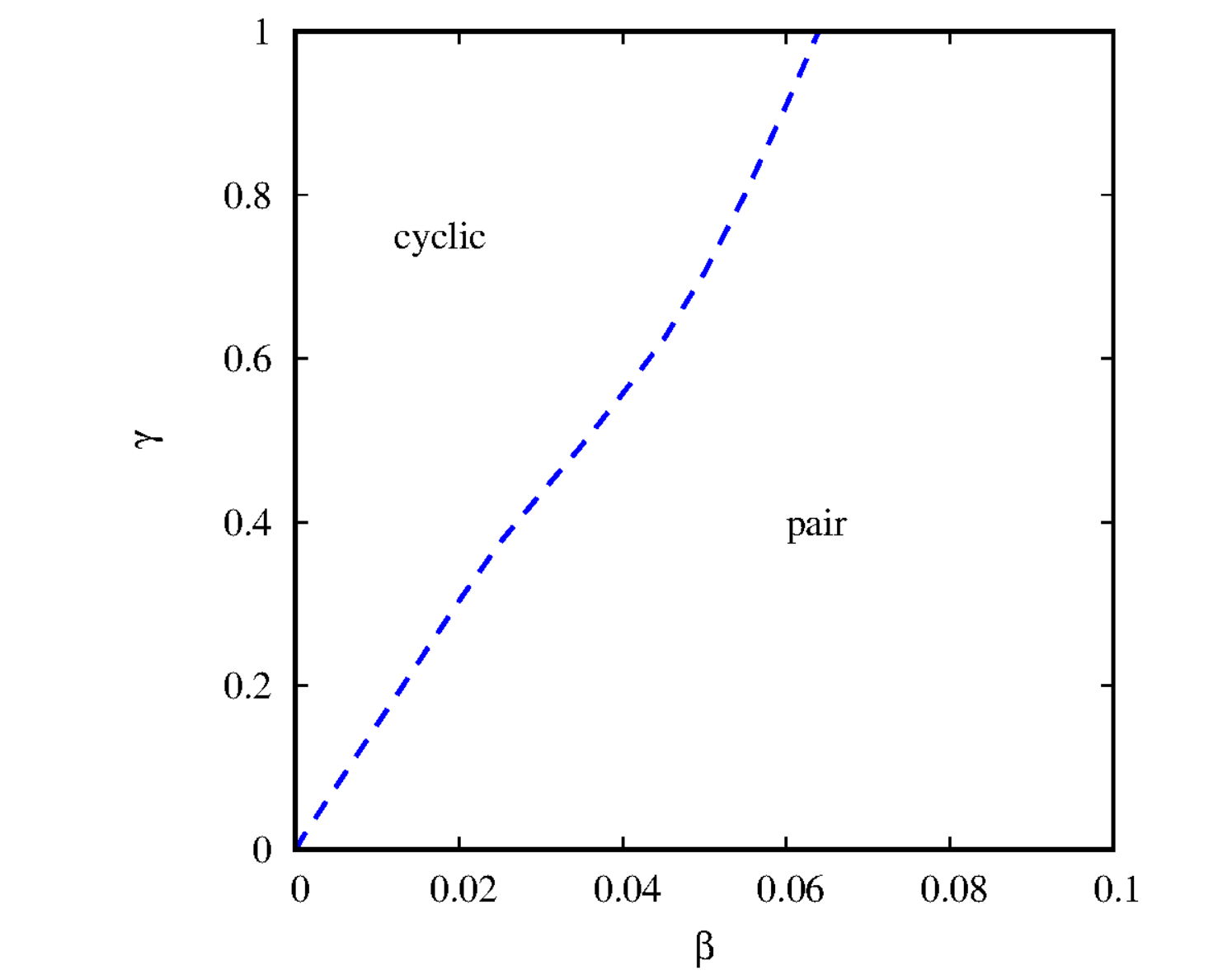}\includegraphics[width=9cm]{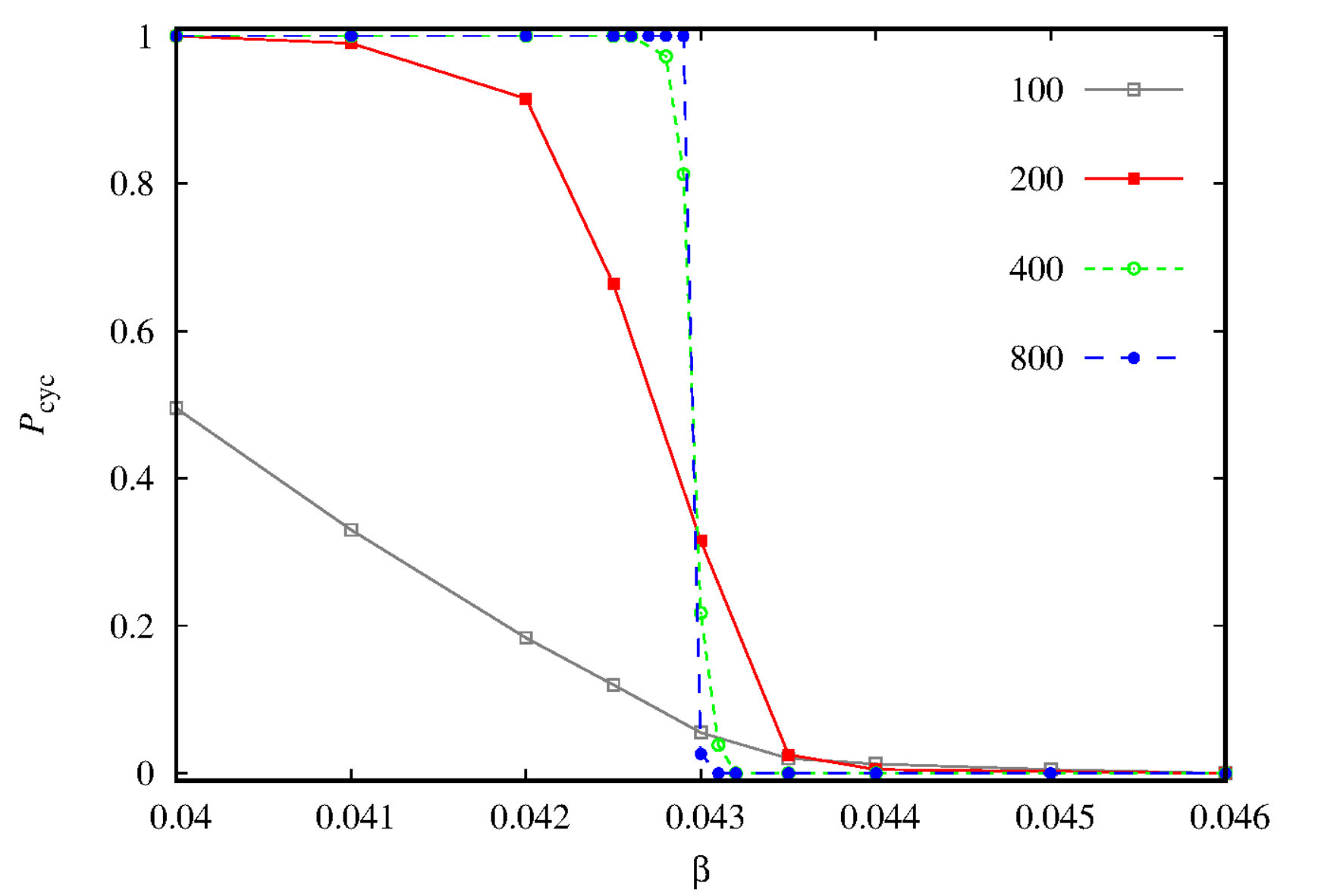}
	\caption{Left panel: phase diagram obtained at $\alpha=0.9$ when the invasion within the cyclic alliance is intensive. The solutions which dominate the $\beta-\gamma$ parameter plane are separated by a discontinuous phase transition line. Right panel: fixation probability to reach the four-member cyclic solution in dependence of $\beta$ at fixed $\gamma=0.6$. The simulations were launched from a prepared initial state shown in the right panel of Fig.~\ref{fs}. The linear system sizes are shown in the legend.}
	\label{a9}
\end{figure}

The quantitative differences of presented phase diagrams motivated us to check how system behavior changes if we gradually change the interaction strength between the alliances. First, we show a representative phase diagram obtained at a large $\alpha=0.9$ value. Here, the inner invasion among the members of quartet is intensive, which makes their defensive alliance effective. As the left panel of Fig.~\ref{a9} indicates, the presence of the strong cyclic solution makes the diagram simpler: either the mentioned alliance or the rival group of pair wins, but there is no chance for the six-member solution to emerge. This conclusion is in harmony with our previous observations because the latter solution could only survive if none of the competing alliances was strong enough. The new diagram also confirms our previous finding about how the interaction strength alters the relation of rival alliances. Namely, as we increase $\gamma$, the cyclic solution becomes even stronger and can compensate the pair solution for relatively large $\beta$ values. The opposite is also true: as we lower $\gamma$, hence the interaction between the competing alliances is weakened, the pair solution becomes dominant even at very small $\beta$ values.

Similarly to previous cases, the usage of prepared initial is proved to be very useful. To illustrate it, we present the fixation probability to reach the cyclic solution in dependence of $\beta$ at a fixed $\gamma=0.6$ value. As the right panel of Fig.~\ref{a9} illustrates, here already the $L=400$ linear system size provides acceptable accuracy, but the application of $L=800$ system size makes our prediction precise by yielding the critical mixing rate $\beta_c = 0.04295 \pm 0.00005$.

\begin{figure}[h!]
	\centering
	\includegraphics[width=7.7cm]{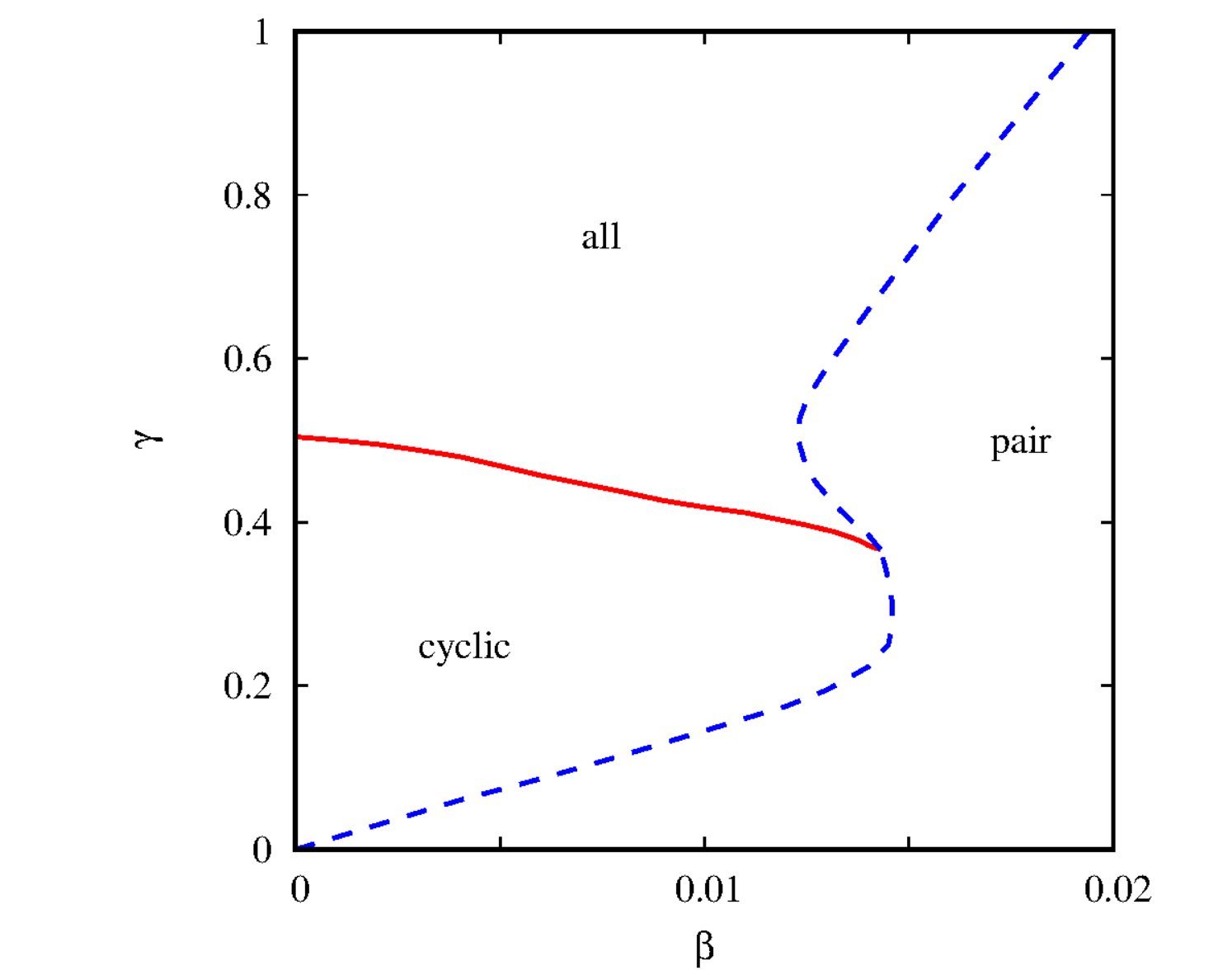}\includegraphics[width=9cm]{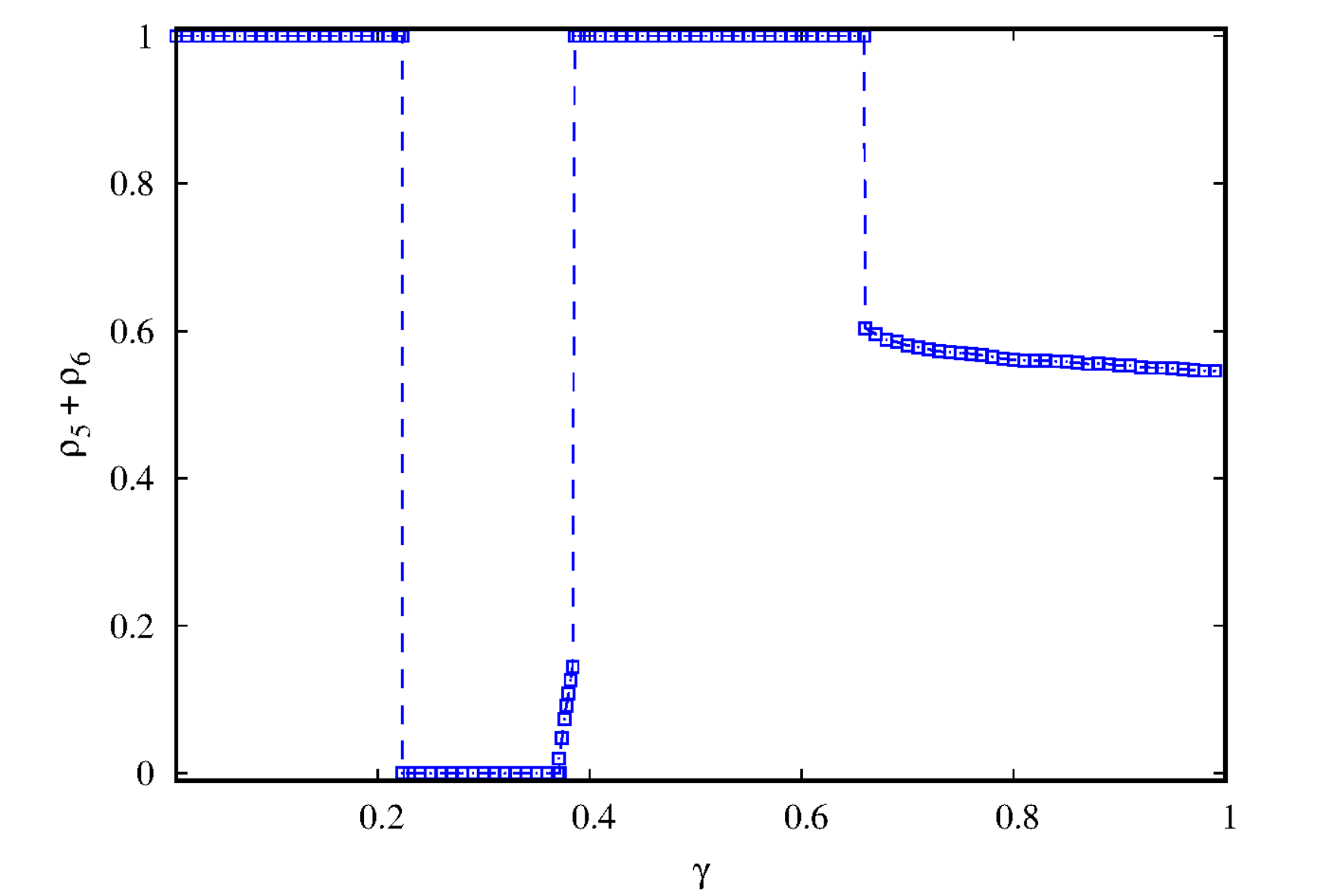}
	\caption{Left panel: phase diagram obtained at $\alpha=0.2$ when the invasion within the cyclic alliance is moderate. Blue dashed (red solid) line represents discontinuous (continuous) phase transitions. Right panel: Cross section of the diagram obtained at $\beta=0.014$. As we increase the value of $\gamma$, the change of the $\rho_5+\rho_6$ order parameter marks four consecutive transition between ``pair'' $\to$ ''cyclic'' $\to$ ''all'' $\to$ ``pair'' $\to$ ``all'' phases.}
	\label{a2}
\end{figure}

To complete our survey of the parameter space, we also explore the phase diagram for a small $\alpha$ value when the cyclic alliance is weak. Our results are summarized in the left panel of Fig.~\ref{a2}. If we compare this diagram with the previous one then several interesting conclusions can be made. First, we would like to note that the horizontal range of the phase diagram is just one-fifth of the range presented in Fig.~\ref{a9}. It simply means that the pair solution becomes already dominant at very small mixing rate. Previously we concluded that large $\gamma$ values help the cyclic solution. But the latter is weak now, hence the up-left corner of $\beta-\gamma$ parameter plane is occupied by the new solution where all six species are present. In general, the decrease of $\gamma$ should support pair solution, but the mixing rate is extremely small now. Therefore, the cyclic quartet can prevail in the down-left region because the uniform $\alpha$ invasion rate makes them stronger against the trios in the new four-member solutions where unequal $\alpha$ and $\gamma$ invasion rates are present in the loop \cite{szolnoki_epl20}. This is again a nice example to illustrate that a larger loop can be fitter if it uses homogeneous invasion rates.

One of the key messages of our work could be that new solutions or subtle system behavior can only be expected in the parameter region where none of the competing alliances are strong enough to dominate the evolution. To illustrate it, we present a cross-section of the last phase diagram at a fixed $\beta=0.014$ value. The $\rho_5+\rho_6$ order parameter is shown in the right panel of Fig.~\ref{a2}. Here we can detect four(!) consecutive phase transitions by only changing the interaction strength between the alliances. Starting from the ``pair'' solution at small $\gamma$ we reach the ``cyclic'' solution via a discontinuous phase transition first. By increasing $\gamma$ further we gradually enter into the six-species phase, which is followed by the ``pair'' solution via a sudden change. If we increase $\gamma$ even further then we can reach the ``all'' phase again. In this way we can detect two reentrant transitions for two different (``pair'' and ``all'') phases by changing a single control parameter.

\section*{Discussion}

Our principal goal was to identify those features which may determine the fitness of a defensive alliance when it fights against a rival group. Similar research question was raised previously \cite{perc_pre07b,de-oliveira_csf22}, but our present work focused on a case when two groups with unequal sizes compete in a symmetrical way. We stress that the defensive mechanisms how a group protects their members from an external invader are different. While in the smaller unit the actors exchange their positions with a certain rate, the members of the quartet may invade each other cyclically. Importantly, we assume that the members of these two groups can invade each other in a balanced way, which results in a three-parameter model. To answer the original question we systematically explored the whole parameter space. 

We found that the two-member alliance is extremely effective. They practically rule the parameter space independently of the other parameters. The rival alliance has only chance to win if the mixing rate between the pair is very low. Importantly, the cyclic solution needs an intensive inner invasion flow to win. This observation confirms previous findings obtained from the competition of three-member loops. Interestingly, the cyclic solution has a better chance of winning if the interaction between the alliances is intensive. In the other extreme case, when members of different groups hardly invade each other, the site exchange becomes even more dominant. The more subtle system behavior can be identified in the parameter region where none of the alliances is strong. This happens if the parameters, which determine the site exchange and the inner invasion, are small. In this case new four-member solutions emerge which are based on a three-member loop extended by the external member of the pair. According to the defensive alliance principle, which is valid for a three-member loop, the external species should be defeated, but the site exchange with the other partner prevents the original mechanism to work, hence establishing a four-member solution. We can identify four similar solutions which interact via the original invasion loop of the cyclic alliance. In this way they form a delicate balance, hence all six species can coexist stably.

When we searched for the stable solution at specific parameter values, we have detected serious finite-size effects, especially in the case if the evolution was launched from random initial state. But this difficulties can practically be avoided if we apply prepared initial states where the competing solutions can fight from the very beginning.

It is worth stressing that our abstract model cannot be applied directly to a real-life system, but the mechanism and principles we revealed could be the basic elements to understand more complex system behaviors where many species are fighting. Furthermore, for simplicity, we used the term ``species'' to describe the actors of this system, but such kind of interactions are not restricted to biological systems \cite{cazaubiel_jtb17,han_xz_amc16,palombi_epjb20}. There are other evolutionary game models, mostly motivated by human society, where competing strategies behave in a similar way \cite{szolnoki_pre10b,griffin_pre17,canova_jsp18,liu_lj_rsif22}.

\section*{Acknowledgements}
A.S. was supported by the National Research, Development and Innovation Office (NKFIH) under Grant No.~K142948. X.C. was supported by the National Natural Science Foundation of China (Grant Nos. 61976048 and 62036002).

\section*{Author contributions statement}
A. S. and C. X. designed and performed the research as well as wrote the paper.

\section*{Additional information}
\textbf{Competing interests} The authors declare no competing interests.

\section*{Data Availability}
 All data generated or analysed during this study are included in this published article.

\end{document}